\documentclass[pre,aps,twocolumn,showpacs]{revtex4}

\usepackage{dcolumn}
\usepackage{amssymb}
\usepackage{graphicx}
\usepackage{epsf,epsfig}

\newcommand \be {\begin{equation}}
\newcommand \ee {\end{equation}}
\newcommand \bea {\begin{eqnarray}}
\newcommand \eea {\end{eqnarray}}

\newcommand \vp {\varphi}

\newfont{\sy}{cmsy10}

\begin{document}

\title{Intensive thermodynamic parameters in nonequilibrium systems}

\author{Eric Bertin$^{1,2}$, Kirsten Martens$^1$, Olivier Dauchot$^3$
and Michel Droz$^1$}
\affiliation{$^1$ Department of Theoretical Physics, University of Geneva,
CH-1211 Geneva 4, Switzerland\\
$^2$ Laboratoire de Physique de l'ENS Lyon, CNRS UMR 5672, 46 All\'ee d'Italie, F-69364 Lyon Cedex 07,
France\\
$^3$ SPEC, CEA Saclay, F-91191 Gif-sur-Yvette Cedex, France}
\date{\today}

\begin{abstract}
Considering a broad class of steady-state nonequilibrium systems
for which some additive quantities are conserved by the dynamics,
we introduce from a statistical approach intensive thermodynamic
parameters (ITPs) conjugated to the conserved quantities.
This definition does not require any detailed balance relation to be
fulfilled. Rather, the system has to satisfy a general additivity property,
which holds in most of the models usually considered in the literature,
including those described by a matrix product ansatz with finite matrices.
The main property of these ITPs is to take equal values in two subsystems,
making them a powerful tool to describe nonequilibrium phase coexistence,
as illustrated on different models.
We finally discuss the issue of the equalization of ITPs when two different
systems are put into contact. This issue is closely related to the possibility
of measuring the ITPs using a small auxiliary system, in the same way as
temperature is measured with a thermometer, and points at one of the major
difficulties of nonequilibrium statistical mechanics.
In addition, an efficient alternative determination, based on the measure
of fluctuations, is also proposed and illustrated.
\end{abstract}


\maketitle

\section{Introduction}

The general concept of intensive thermodynamic parameters plays a 
crucial role in equilibrium statistical mechanics.
For systems in contact, ITPs like temperature, 
pressure or chemical potential equalize their values once the 
equilibrium state is reached, provided that their associated 
quantities, energy, volume or number of particles, can be exchanged.
Since this property holds even in the case of systems
that exhibit different microscopic dynamics, this equalization 
became the key criterion in equilibrium statistical mechanics to study 
the influence of the environment on a given system, for example when
a reservoir is connected to it.
Moreover the theory of phase coexistence 
as well as the measurement of, for instance, temperature with a thermometer
draw on this powerful concept.

Indeed this potency of the ITP formalism motivates the endeavor to 
generalize the notion of ITPs to nonequilibrium systems. There exist
several different approaches mostly focusing on the generalization of 
temperature out of equilibrium.
For stationary nonequilibrium systems that fulfill 
a local equilibrium condition, equilibrium properties are locally 
recovered, so that ITPs can be naturally defined on macroscopic scales that
remain small in comparison with the system size \cite{deGroot}.
Beyond local equilibrium, more phenomenological 
endeavors based on thermodynamical grounds \cite{Jou} have been proposed,
as well as statistical approaches illustrated on some specific models
\cite{Puglisi,MaxEnt,Hatano,BDD,Jou06}.
Finally, in the case of non-stationary slow 
dynamics, a notion of temperature may be derived from a 
generalized fluctuation-dissipation relation (FDR) \cite{CuKuPe, 
Kurchan, Crisanti, Ritort} in analogy to equilibrium statistical mechanics.
Still, in spite of these numerous propositions, the relevance of this concept
of effective temperature and its possible generality have been 
barely discussed.

Besides, other notions of ITPs have appeared in the recent literature
on nonequilibrium systems. For instance, in the context of stochastic
models with a conserved mass $M$ (or number of particles),
like the Zero Range Process or different
kinds of mass transport models \cite{Evans-Rev05},
a formal grand-canonical ensemble has been defined,
in which systems with a total mass $M$ appear with a
probability weight proportional to $\exp(-\mu M)$, $\mu$ being called
a chemical potential.
We call this ensemble a ``formal'' one, since no definition of the chemical
potential is given prior to the grand-canonical construction, and no
physical mechanism allowing for fluctuations of the total mass
(like, e.g., a contact with a reservoir) is described.
Hence, the grand-canonical distributions introduced so-far appear
more as mathematical tools with interesting properties,
as it may be considered as a Laplace transform with respect
to $M$ of the canonical distribution for which $M$ is fixed
\cite{Arndt,Evans-Rev05}.
Even more importantly, if this formal grand-canonical ensemble was
to be considered as defining the chemical potential, nothing could
be said from it on the possible equalization of this parameter
between two subsystems of a globally isolated system, since in the
grand-canonical ensemble, the chemical potential is externally
imposed.

The aim of the present work is to introduce a precise and general theoretical
background allowing for the definition of ITPs conjugated
to conserved quantities in nonequilibrium systems.
Note that the systems considered here are out of equilibrium not due to
the presence of gradients imposed, for instance, by boundary reservoirs,
but because of the breaking of microreversibility (that is, time-reversal
invariance) at the level of the microscopic dynamics in the bulk.
Accordingly, the ITPs are \emph{not} space dependent, as would be
the case for systems that fulfill the local equilibrium assumption.
We state the hypotheses underlying the present construction, and clarify
the physical interpretation of the grand-canonical ensemble.
We then discuss the relevance and usefulness of the concept of ITP, with
particular emphasis on the description of phase coexistence.
Although the proposed generalization of ITPs
appears to be rather natural, it turns out that non-trivial problems arise
as soon as systems with different dynamics are put into contact.
This issue is essential if one wants to measure an ITP using a small
auxiliary system --just like temperature is measured with a thermometer--
and points at one of the main difficulties of nonequilibrium statistical
mechanics. Note that part of the results reported here have appeared
in a short version \cite{short}.

The paper is organized as follows. Sec.~\ref{sec-def} introduces
the concept of intensive thermodynamic parameters in the framework
of nonequilibrium systems. The main condition 
of validity for this concept is given and some features in analogy to 
equilibrium statistical physics are discussed. Further
the applicability of the definition for systems described with
matrices (either matrix product ansatz or transfer matrix method)
is studied, and an illustration
of this approach on two simple models is given.
Sec.~\ref{sec-phase-co} is dedicated to the issue of phase separation,
specifically the problem of condensation, in different models. We show
how the concept of ITPs could improve the understanding
of phase separation in these kind of models.
Finally, the contact of two systems with different dynamics is
investigated in Sec.~\ref{sec-contact}, with emphasis on the
issue of equalization of ITPs. We also explore possible ways 
of determining the ITPs in experiments or numerical simulations.

\section{Nonequilibrium intensive thermodynamic parameters}
\label{sec-def}

\subsection{Framework and definitions}
\label{sec-frame-def}

Let us start by
considering a general macroscopic system that exhibits a steady state, and
such that the dynamics conserves some additive quantities, referred to
as $Q_k$, $k=1,\ldots,\ell$, in the following.
Such systems have been extensively studied for instance
in the context of markovian stochastic models, and simple examples include
the zero-range process (ZRP) \cite{Evans-Rev05},
more general mass transport models \cite{Zia04,Zia05,Evans06}, or
the asymmetric exclusion process (ASEP) in a closed geometry \cite{Arndt}.
On the microscopic level, the nonequilibrium character of the dynamics
manifests itself (apart from the lack of detailed balance)
in the presence of a non-trivial dynamical weight $f_{\alpha}$ associated with
each microscopic configuration $\alpha$, in the
steady-state probability $P_{\alpha}$. More precisely, the latter reads 
\begin{equation}\label{dist-mucan}
P_{\alpha} = \frac{f_{\alpha}}{Z(Q_1,\ldots,Q_{\ell})}
\prod_{k=1}^{\ell} \delta(Q_k^{\alpha}-Q_k)\, ,
\end{equation}
where the product of delta distributions ensures the conservation of
the quantities $Q_k$. The function
\begin{equation}
Z(Q_1,\ldots,Q_{\ell})=\sum_{\alpha}f_{\alpha}
\prod_{k=1}^{\ell} \delta(Q_k^{\alpha}-Q_k)
\end{equation}
serves as normalization factor, which will be referred to as ``partition
function'' in analogy to equilibrium statistical mechanics.
Let us emphasize that in equilibrium systems, the probability weight
$f_{\alpha}$
is either a constant independent of $\alpha$, or an exponential factor
accounting for the exchange of a conserved quantity with a reservoir.
For instance, in the equilibrium canonical ensemble, the conserved quantity
would be the number of particles, whereas $f_{\alpha}$ would be the Gibbs
factor $\exp(-\beta E_{\alpha})$, where $E_{\alpha}$ is the energy of
configuration $\alpha$, and $\beta$ is the inverse temperature.
On the contrary, in a nonequilibrium system, the weights $f_{\alpha}$ also account
for purely dynamical effects related to the absence of microreversibility (the latter
being deeply rooted in the hamiltonian properties of equilibrium systems), so that
these weights generically depart from a constant, even if no conserved quantity is
exchanged with a reservoir.

To introduce a definition for ITPs in nonequilibrium situations, we first
recall that their equilibrium definition is related to the exchanges
of conserved quantities between subsystems, and ensures the equality of
ITPs in different parts of the system.
Following the same line of thought in a nonequilibrium context, let us divide
our system, in an arbitrary way, into two subsystems $\mathcal{S}_a$ and $\mathcal{S}_b$. The sum $Q_{ka}$+$Q_{kb}=Q_k$ is kept
constant due to the conservation law, whereas exchanges of these quantities
between the two subsystems are allowed.
The microstate $\alpha$ is now defined as the combination of the
two microstates $\{\alpha_a,\alpha_b\}$ of the subsystems,
so that the probability
of a microstate is denoted as $P_{\alpha_a,\alpha_b}$.
An important quantity in the following approach is the conditional probability
$\Psi(Q_{1a},\ldots,Q_{\ell a}|Q_1,\ldots,Q_{\ell})$ that
the conserved quantities have values $Q_{ka}$ in subsystem $\mathcal{S}_a$,
given their total values $Q_k$
\begin{equation}
\Psi(\{Q_{ka}\}|\{Q_k\}) =
\sum_{\alpha_a,\alpha_b} P_{\alpha_a,\alpha_b}
\prod_{k=1}^{\ell} \delta(Q_{ka}^{\alpha_a}-Q_{ka})\, .
\end{equation}
The key assumption in the following derivation is that the logarithm of
$\Psi(\{Q_{ka}\}|\{Q_k\})$ satisfies an asymptotic additivity property,
namely
\begin{eqnarray} \label{afp}
\ln \Psi(\{Q_{ka}\}|\{Q_k\}) &=& 
\ln Z_a(\{Q_{ka}\}) + \ln Z_b(\{Q_k-Q_{ka}\})\nonumber\\ 
&-& \ln Z(\{Q_k\}) + \epsilon_N(\{Q_{ka}\},\{Q_k\})
\end{eqnarray}
with
\begin{equation} \label{afp2}
|\epsilon_N(\{Q_{ka}\},\{Q_k\})|\ll|\ln \Psi(\{Q_{ka}\}|\{Q_k\})|\; ,
\end{equation}
in the thermodynamic limit $N \to \infty$.
In Eq.~(\ref{afp}), $Z_{\nu}(\{Q_{k\nu}\})$ refers to the isolated subsystem
$\mathcal{S}_{\nu}$, $\nu\in\{a,b\}$, and N is the number of degrees of
freedom.
That this additivity condition is fulfilled for some rather large classes
of nonequilibrium systems will be illustrated in the following examples.

Some of the simplest systems that fulfill Eqs.~(\ref{afp}), (\ref{afp2})
are lattice models with a factorized steady state distribution
(where sites are labelled by $i=1, \dots, N$, and $\alpha=\{\alpha_i\}$)
\begin{equation}
P_{\alpha}= \frac{1}{Z(Q_1,\ldots,Q_{\ell})}\prod_{i=1}^N f_{i,\alpha_i}
\prod_{k=1}^l \delta\left(\sum_{i=1}^N Q_{ki}^{\alpha_i}-Q_k \right)\;,
\end{equation}
in which case the term $\epsilon_N$ vanishes. 
Well-known examples of models with factorized steady-states
are for instance the ZRP
\cite{Evans-Rev05} and other general mass transport models \cite{Zia04, Zia05}.
Accordingly, the physical interpretation of the additivity condition
given in Eqs.~(\ref{afp}), (\ref{afp2}) is that, on large scale, the system
behaves essentially as if the probability weight was factorized,
although the genuine probability weight may not be factorized. 

As mentioned above, our aim is to define a parameter that takes equal values
within the two (arbitrary) subsystems $\mathcal{S}_a$ and $\mathcal{S}_b$.
Guided by the equilibrium procedure, we consider the
most probable value of $Q_{ka}$, denoted by $Q_{ka}^*$, which maximizes the probability
$\Psi(\{Q_{ka}\}|\{Q_k\})$. This most probable value satisfies
\begin{equation}
\frac{\partial \ln \Psi(\{Q_{ka}\}|\{Q_k\}) }{\partial Q_{ka}}\Big\vert_{Q_{ka}^*}=0\, .
\end{equation}
Using the additivity condition (\ref{afp}), (\ref{afp2}), we obtain the following relation
\begin{equation}\label{equil-ITP}
\frac{\partial \ln Z_a}{\partial Q_{ka}} \Big\vert_{Q_{ka}^*} =
\frac{\partial \ln Z_b}{\partial Q_{kb}} \Big\vert_{Q_k-Q_{ka}^*}.
\end{equation}
Hence, it is natural to define the ITP $\lambda_k$ conjugated to the
conserved quantity $Q_k$ in a nonequilibrium system as 
\begin{equation}\label{def-ITP}
\lambda_k \equiv \frac{\partial \ln Z}{\partial Q_k}\;,
\qquad k=1,\dots, \ell\;.
\end{equation}
Once the steady state is reached, this quantity equalizes in the
two subsystems $\mathcal{S}_a$ and $\mathcal{S}_b$ due to
Eq.~(\ref{equil-ITP}), and thus satisfies the basic requirement for the
definition of an ITP.

Actually, for the approach to be fully consistent, one has to check 
that the value of $\lambda_k$ does not depend on the choice of
the partition, as long as both subsystems remain macroscopic.
We show in Appendix~\ref{app-choice-indep} that this is indeed the case,
at least under the assumption (which is consistent with the additivity
condition) that $\ln Z$ is extensive.
To achieve this result one shows that
the ITP $\lambda_k$ obtained from the subsystems is equal to the global ITP
defined on the whole system from Eq.~(\ref{def-ITP}), independently
of the partition chosen (see Appendix \ref{app-choice-indep}). 
Thus from now on, we compute the ITP $\lambda_k$ using Eq.~(\ref{def-ITP})
for the whole system.

\subsection{Nonequilibrium ``grand-canonical'' ensemble}

Now that we have defined the notion of ITP, it is natural to try to
introduce a grand-canonical ensemble describing a system in contact
with a reservoir of conserved quantities, that imposes its values of
$\{\lambda_k\}$ to the system. Let us consider a partition
of a large isolated system into two subsystems
$\mathcal{S}_a$ and $\mathcal{S}_b$, such that
one of the macroscopic systems is much smaller than the other.
The larger subsystem, say $\mathcal{S}_b$, serves as a reservoir for the 
$\{Q_k\}$, leading to a natural definition of a ``grand-canonical'' distribution.
This distribution is obtained by integrating the ``canonical''
distribution~(\ref{dist-mucan}) over the degrees of freedom of the reservoir
\begin{eqnarray}\label{dist-can}
\tilde{P}_{\alpha_a} &=& \frac{f_{\alpha_a}}{Z(Q_1,\ldots,Q_{\ell})}
\sum_{\alpha_b} f_{\alpha_b}\prod_{k=1}^{\ell} \delta(Q_{ka}^{\alpha_a}
+Q_{kb}^{\alpha_b}-Q_k)\nonumber\\
&=& \frac{f_{\alpha_a}}{Z(Q_1,\ldots,Q_{\ell})} Z_b(\{Q_k-Q_{ka}^{\alpha_a}\})
\end{eqnarray}
Note that in the above equation, we assume the factorization property
$f_{\alpha_a,\alpha_b} = f_{\alpha_a} f_{\alpha_b}$, that is we neglect
possible boundary contributions, similarly to the assumptions often made
in the equilibrium context.
Expanding the logarithm of the partition function $Z_b$ for 
$Q_{ka}^{\alpha_a}\ll Q_k$, one finds to leading order in $Q_{ka}^{\alpha_a}$
\begin{eqnarray}
\ln Z_b(\{Q_k-Q_{ka}^{\alpha_a}\})&\approx &\ln Z_b(\{Q_k\})-
\sum_{k=1}^{\ell} \frac{\partial \ln Z_b}{\partial Q_{kb}} \Big\vert_{Q_k}
Q_{ka}^{\alpha_a}
\nonumber\\
&\approx & \ln Z_b(\{Q_k\})-\sum_{k=1}^{\ell} \lambda_k Q_{ka}^{\alpha_a}\;,
\end{eqnarray}
where we approximated
\begin{equation}
\frac{\partial \ln Z_b}{\partial Q_{kb}} \Big\vert_{Q_k}
\approx \frac{\partial \ln Z_b}{\partial Q_{kb}} \Big\vert_{Q_{kb}^*}
=\lambda_k\;.
\end{equation}
This is justified in the limit $N_a/N\to 0$. Therefore we obtain for the ``grand-canonical''
distribution the following expression
\begin{equation}
\tilde{P}_{\alpha_a}=\frac{f_{\alpha_a}}{\tilde{Z}(\lambda_1,\ldots,
\lambda_{\ell})}\, \exp\left(-\sum_{k=1}^l\lambda_kQ_{ka}^{\alpha_a}\right)\;.
\label{dist-grand-can}
\end{equation}
which defines the grand-canonical partition function 
\be
\tilde{Z}(\lambda_1,\ldots,\lambda_{\ell}) =
\sum_{\alpha_a} f_{\alpha_a} \, \exp\left(-\sum_{k=1}^l\lambda_kQ_{ka}^{\alpha_a}\right)\;.
\ee
Interestingly, the cumulants $<Q_{k_1} Q_{k_2} \cdots Q_{k_n}>_c$ may be 
expressed
as a derivative of the logarithm of the grand-canonical partition function
$\tilde{Z}(\{\lambda_k \})$, which thus appears as the associated
generating function, as follows:
\begin{eqnarray}\label{cum}
<Q_{k_1} Q_{k_2} \cdots Q_{k_n}>_c &=& (-1)^n \\ \nonumber
&&\frac{\partial }{\partial \lambda_{k_1}}
\frac{\partial }{\partial \lambda_{k_2}} \cdots \frac{\partial }{\partial
\lambda_{k_n}} \ln \tilde{Z}(\{\lambda_k\})
\end{eqnarray}
Note that some of the indices among $k_1,\ldots,k_n$ may be the
same. This result generalizes the corresponding well-known equilibrium
result \cite{Fulde}. Accordingly, checking the validity of such relations
in a given system does not show that microstates compatible with the
constraints are equiprobable, contrary to what is sometimes implicitely
assumed (e.g., in the context of granular matter \cite{Swinney}).

\subsection{Applicability for nonfactorized steady states}

As stated above every nonequilibrium system that exhibits a product measure 
automatically fulfills the additivity property given in
Eqs.~(\ref{afp}), (\ref{afp2}), since $\epsilon_N$ vanishes in this case.
The aim of this section is to discuss some typical cases
for which the steady-state distribution does not factorize,
but still satisfies the additivity condition.
As a result, the ITP framework is relevant for such systems.
There are two important classes of models with nonfactorized
steady states:
systems where the stationary state can be expressed by a matrix product
ansatz, and models described by a transfer matrix.

\subsubsection{Matrix product ansatz}

Practically speaking, considering a matrix product ansatz means that,
in a one-dimensional system with periodic boundary conditions,
one expresses the probability weight $f_{\alpha}$ in the form
\be
f_{\alpha} = \mathrm{Tr} \prod_{i=1}^N M_{\alpha_i}
\ee
where $M_{\alpha_i}$ is a (possibly infinite) matrix, $\alpha_i$ is the
state of site $i$, and $\mathrm{Tr}$ is the trace operation over
the matrices.
For nonperiodic boundary conditions,
a slightly different ansatz is used, namely
\be
f_{\alpha} = \langle W|\prod_{i=1}^N M_{\alpha_i}|V \rangle
\ee
where the vectors $\langle W|$ and $|V \rangle$ are determined by the
boundary conditions (reflecting boundaries or injection of particles from
a reservoir for instance). Matrix product ansatz have proved particularly
useful in the context of the ASEP,
where particles obeying an exclusion principle (that is,
at most one particle per site is allowed) perform a biased stochastic
motion on a one-dimensional lattice \cite{Derrida93,Mallick,Arndt}.

To test the additivity condition (\ref{afp}), (\ref{afp2}) for such models,
let us consider a generic lattice model with periodic boundary
conditions. A variable $q_i$ is defined on each site $i=1,\dots N$, and we
assume that the quantity $Q=\sum_{i=1}^N q_i$ is conserved by the dynamics.
The steady-state distribution is assumed to be described by a
matrix product ansatz:
\be \label{model-MPA}
P(\{q_i\}) = \frac{1}{Z(Q)} \, \mathrm{Tr} \prod_{i=1}^N M(q_i) \,
\delta\left(\sum_{i=1}^N q_i -Q\right),
\ee
where $M(q)$ is a square matrix. Let us also introduce
the matrix $R(Q)$ through
\be
R(Q) = \int \prod_{i=1}^N [dq_i\; M(q_i)] \,
\delta\left(\sum_{i=1}^N q_i -Q\right),
\ee
so that the normalizing factor $Z(Q)=\mathrm{Tr}\, R(Q)$.
This leads for the conditional probability distribution $\Psi(Q_a|Q)$ to:
\be \label{psi-MPA}
\Psi(Q_a|Q) = \frac{1}{Z(Q)} \, \mathrm{Tr} [R_a(Q_a) R_b(Q-Q_a)]
\ee
Loosely speaking, the additivity condition holds if the last factor
$\mathrm{Tr} [R_a(Q_a) R_b(Q-Q_a)]$ behaves essentially as
$Z_a(Q_a) Z_b(Q-Q_a)$.
It is shown in Appendix~\ref{app-matrix} that the additivity property
(\ref{afp}), (\ref{afp2}) is generically fulfilled for
a system described by a matrix product ansatz with
\emph{finite} matrices. Whether it also holds for some classes of
infinite matrices $M(q)$ remains an open issue.

\subsubsection{Models with transfer matrices}

Another matrix method that has become very popular, in equilibrium
as well as in nonequilibrium statistical physics, is the transfer matrix
one.
The main idea of this method is to formulate the 
partition function in terms of a product of a matrix, 
the so-called transfer matrix.

To give an example of the application of this method for nonfactorized 
steady states let us consider, as previously, a one-dimensional transport
model on a ring,
with a local variable $q_i$ on each site $i$, and such that
the sum $Q=\sum_{i=1}^N q_i$ is conserved. The variables $q_i$ may either
be discrete or continuous.
Let us now assume a steady-state distribution of the form
\begin{equation} \label{dist-transf-mat}
P(\{q_i\})=\frac{1}{Z(Q)}\prod_{i=1}^N g(q_i,q_{i+1})\, \delta\left(\sum_{i=1}^N q_i-Q\right)
\end{equation}
with $q_{N+1} \equiv q_1$, and $g(q_i,q_{i+1})$ is a symmetric function.
The partition function $Z(Q)$ is given by
\begin{equation}
Z(Q)=\int \prod_{i=1}^N [dq_i \, g(q_i,q_{i+1})] \, \delta\left(\sum_{i=1}^N
q_i-Q\right)\; ,
\end{equation}
Note that a model of this type has been studied in \cite{Evans06}, showing
interesting nonequilibrium condensation properties.
Let us introduce the quantity $S_N(Q,q_1,q_{N+1})$ defined as
\bea
S_N(Q,q_1,q_{N+1}) &\equiv& \int dq_2 \dots dq_N \\ \nonumber
&\times& \prod_{i=1}^N g(q_i,q_{i+1})
\, \delta\left(\sum_{i=1}^N q_i-Q\right)
\eea
where $q_1$ is no longer identified with $q_{N+1}$, contrary to
Eq.~(\ref{dist-transf-mat}). Then one has
\be
Z(Q) = \int_0^{\infty} dq_1 \, S_N(Q,q_1,q_1) \; ,
\ee
and the distribution $\psi(Q_a|Q)$ can be written as
\bea
\Psi(Q_a|Q) &=& \frac{1}{Z(Q)} \, \int_0^{\infty} dq_1 \int_0^{\infty}
dq_{N_a+1} \\ \nonumber
&\times& S_{N_a}(Q_a,q_1,q_{N_a+1})\, S_{N_b}(Q-Q_a,q_{N_a+1},q_1)
\eea
>From this expression, a calculation similar in spirit to the one presented
in Appendix~\ref{app-matrix} for the case of matrix product ansatz,
allows one to show that the additivity condition
(\ref{afp}), (\ref{afp2}) holds. The derivation makes use of the Laplace transform
$\hat{S}_N(s,q_1,q_{N+1})$ of $S_N(Q,q_1,q_{N+1})$ with respect to $Q$,
which can be written as a matrix product (hence the name transfer matrix
method):
\be
\hat{S}_N(s,q_1,q_{N+1}) = e^{-s(q_1-q_{N+1})/2}\, \mathbf{T}_s^N(q_1,q_{N+1})
\ee
where the transfer matrix $\mathbf{T}_s$ is defined by
\be
\mathbf{T}_s(q,q') = g(q,q') \, e^{-s(q+q')/2}
\ee
Let us mention, here again, that the derivation of the additivity condition
relies on some properties of the transfer matrix that are well established
for finite matrices, but that might not be fulfilled in some cases
for infinite matrices.

\subsection{ITPs at work on simple models}
\label{homo-model}

\subsubsection{Mass transport model with factorized steady-state}
\label{fact}
As a first example we consider a simple one-dimensional mass transport
model on a ring with only one globally conserved quantity, referred to as
mass $M=\sum_{i=1}^Nm_i$, as introduced in Ref.~\cite{Zia04}.
The masses $m_i$ are a priori positive and real variables.
The continuous time stochastic dynamics is defined as follows.
A mass $\mu$ is transferred from a randomly chosen
site $i$, containing the mass $m_i$, to site $i+1$ according to the
following rate
\begin{equation}\label{mtm-rate}
\varphi_i(\mu|m_i)=v(\mu)\frac{f_i(m_i-\mu)}{f_i(m_i)}\;.
\end{equation}
Thus transport is totally biased, which generates a flux of mass
along the ring.
With the above rate, the steady-state distribution is of the form
\cite{Zia04} (see also Appendix~\ref{app-pair} for a more general case)
\begin{equation}
P(\{m_i\})=\frac{1}{Z(M)} \prod_{i=1}^N f_i(m_i) \,
\delta\left(\sum_{i=1}^N m_i-M\right)\;,
\end{equation}
where the single-site weight $f_i(m_i)$ may be site-dependent.
Let us here consider for $f_i(m_i)$ the simple form $f_i(m_i)=m_i^{\eta_i-1}$
with $\eta_i>0$ for all $i$.

To calculate the ITP corresponding to the conserved mass in the system
we need to find the dependence of the partition function $Z$ on M:
\begin{equation}
Z(M)= \int \prod_{i=1}^N [dm_i\, m_i^{\eta_i-1}]\, \delta\left(\sum_{i=1}^N m_i-M\right)\; ,
\end{equation}
where the integrals are over the positive real axis.
A simple rescaling $m_i=x_i M$ reveals the searched dependence:
\begin{eqnarray}
Z(M)&=&M^{\sum_{i=1}^N \eta_i-1} \int \prod_{i=1}^N [dx_i\, x_i^{\eta_i-1}]\, \delta\left(\sum_{i=1}^Nx_i-1\right)\nonumber\\
&=&K_N M^{N\overline{\eta}-1}
\end{eqnarray}
with $\overline{\eta}=N^{-1}\sum_{i=1}^N \eta_i$, and where
$K_N$ is a constant independent of M. 
The ITP is obtained from the derivative of $\ln Z$
\be
\lambda = \frac{d \ln Z}{d M}
= \frac{N\overline{\eta} -1}{M}
\ee
leading in the thermodynamic limit $N \to \infty$ to
\be
\lambda = \frac{\overline{\eta}}{\rho}\;,
\ee
where $\rho=M/N$ denotes the average density.

\subsubsection{Model with pair-factorized steady-state}
\label{pair}

Let us now consider a second example
to illustrate that our approach works as well on
systems that do not exhibit a factorized steady state. We
therefore consider a model similar to that used in the preceding section,
but with a transport rate that depends, besides the mass on
the concerned site, also on the masses on the two neighboring
sites:
\begin{eqnarray} \label{pair-rate}
\varphi(\mu|m_{i-1},m_i,m_{i+1})&=&v(\mu)\frac{g(m_{i-1},m_i-\mu)}{g(m_{i-1},m_i)}\times\nonumber\\
&&\frac{g(m_i-\mu,m_{i+1})}{g(m_i,m_{i+1})}\;.
\end{eqnarray}
Note that this is a generalization of the dynamics introduced in
\cite{Evans06}, defined for discrete masses.
It can be shown (see Appendix~\ref{app-pair}) that these dynamics lead to
a pair factorized steady state of the form
\begin{equation} \label{pair-dist}
P(\{m_i\})=\frac{\prod_{i=1}^N g(m_i,m_{i+1})}{Z(M)}\, \delta\left(\sum_{i=1}^N m_i-M\right)\;,
\end{equation}
Choosing, as a simple example, the function $g(m,n)$ as
\begin{equation} \label{g-abc}
g(m,n)=(m^{\alpha} n^{\beta}+m^{\beta}n^{\alpha})^{\gamma}
\end{equation}
with $\alpha, \beta, \gamma\geq0$, we obtain the behavior of the partition 
function on M, using again a simple scaling argument:
\begin{eqnarray}
Z(M) &=& \int \prod_{i=1}^N [dm_i\, g(m_i,m_{i+1})]\, \delta\left(\sum_{i=1}^Nm_i-M\right)\nonumber\\
&=&M^{N[\gamma(\alpha+\beta)+1]-1} \int \prod_{i=1}^N [dx_i\, g(x_i,x_{i+1})]
\nonumber\\
&& \qquad \times \, \delta\left(\sum_{i=1}^Nx_i-1\right)\nonumber\\
&=& \tilde{K}_N M^{N[\gamma(\alpha+\beta)+1]-1}\;
\end{eqnarray}
with a prefactor $\tilde{K}_N$ independent of M.
Thus the ITP conjugated to the conserved mass reads
\be
\lambda = \frac{d \ln Z}{d M}
= \frac{N[\gamma(\alpha+\beta)+1]-1}{M}
\ee
yielding in the thermodynamic limit
\be
\lambda = \frac{\gamma(\alpha+\beta)+1}{\rho}\;.
\ee
Note that here, we did not checked the additivity condition
(\ref{afp}), (\ref{afp2}), while the distribution (\ref{pair-dist})
would lead to an infinite transfer matrix. Yet, we check in
Sec.~\ref{sub-measure}, on a specific example, that the above results
are indeed consistent.

\subsection{Discussion on the physical interpretation of ITPs}

Now that the formalism is introduced, let us insist on the physical
interpretation of ITPs.
First, it is important to note that, in the present framework, the
grand-canonical ensemble is explicitely derived from the canonical one,
and thereby gains a clear physical status, whereas the ``formal''
grand-canonical ensemble often considered in the literature, is in a
sense more like a Laplace transform, without physical ground \cite{Arndt}.
A nice
illustration of this difference appears when considering the topology of
the physical system which may indeed differ in the canonical and
grand-canonical ensemble. For instance, when dealing with a canonical
system on a one-dimensional ring, the corresponding grand-canonical system
would be on a segment, the complementary segment playing the role of the
reservoir. Such a subtility does not emerge when dealing with the Laplace
transform of the canonical distribution. 
Second, in the present context the ITP has the status of a well-defined
thermodynamical  variable, which equalizes between subsystems.
We shall illustrate in the next section how it applies to the description
of phase coexistence and how it allows to extend our ``thermodynamical way
of thinking'' to some out-of-equilibrium situations.

One might also argue that, at least when the conserved quantity is a number
of particles, the present definition of the associated ITP may be
recovered by mapping the nonequilibrium stationary distribution on an
effective equilibrium canonical distribution (thus introducing an
effective hamiltonian) and computing the corresponding equilibrium
chemical potential. Although it leads to the same result
as ours, such a procedure is less general
and rather confusing, since the nonequilibrium aspects of the system
are somehow hidden. Indeed, one may conclude from such a mapping
that two nonequilibrium systems in contact should equilibrate their
chemical potentials, which is not necessarily true, as will appear more
clearly in Sec.~\ref{sec-contact}.

Besides, it is sometimes thought
that ITPs somehow identify with fluxes within a given system,
for instance the flux of mass between two neighboring sites in a mass
transport model.
One reason for this is that in one-dimensional models, the stationary flux
necessarily takes the same value throughout the system.
Yet, this identification is actually {\it not} valid in general,
but only for special cases such as for instance the
one-dimensional ZRP, where the flux $\Phi$ of
particles within the systems is equal to the fugacity $e^{-\lambda}$.
To illustrate this point,
let us take for example a simple mass transport model on
a ring with $N$ sites and transport rates $\varphi(\mu|m)$ defined
in Eq.~(\ref{mtm-rate}), assuming for simplicity that $f_i(m)=f(m)$
for all $i$.
The average flux $\phi$ crossing a given link reads
\begin{eqnarray}
\phi=\int_0^{\infty} dm\, P(m)\int_0^m d\mu\, \mu \varphi(\mu|m)
\end{eqnarray}
Using the single site probability $P(m)=C f(m) e^{-\lambda m}$, 
C being a normalization constant, one finds
\be
\phi=C\int_0^{\infty} d\mu\, \mu v(\mu) e^{-\lambda \mu}
\int_0^{\infty} dx\, f(x) e^{-\lambda x}
\ee
with the change of variables $x=m-\mu$.
Using the definition of $C$, the above equation reduces to
\be
\phi = \int_0^{\infty} d\mu\, \mu v(\mu) e^{-\lambda \mu}
\ee
Considering this equation we can conclude
that the relation between the flux and the ITP can be highly nontrivial,
and that the two notions should not be identified.
Moreover, there is no obvious reason why, in a generic system of dimension
$d>1$, the local flux should take equal values throughout the system.
Note also that in models with closed boundary conditions,
the global flux of a conserved quantity is forced to be zero,
whereas ITPs a priori take nonzero values, making them a more useful
characterization of non-equilibrium steady states.

\section{Relevance of ITP's to describe phase coexistence}
\label{sec-phase-co}
\subsection{Principle of the approach}

One of the main interest of the notion of ITP, which made its success
in equilibrium, is that such parameters take equal values in different
subsystems of a given system, regardless of their macroscopic state.
Indeed, subsystems may for instance have different densities of a given
conserved quantity,
but they should have the same value of the associated ITP.
A case of great interest where this situation arises is that of phase
coexistence. At equilibrium, phase coexistence is described by the
equality of the different ITPs (temperature, pressure, chemical potential).
We shall now argue that the present ITP formalism allows for a similar
description of nonequilibrium phase coexistence, at least when this
phenomenon is related to a conservation law, and when the additivity
condition (\ref{afp}), (\ref{afp2}) holds.

As a simple illustration, we shall consider in this section a
well-studied example of nonequilibrium phase coexistence, namely
condensation transitions.
Such transitions have been reported in ZRP
\cite{Spohn,Godreche,Hanney,Evans-Rev05}, and in more general mass transport
models \cite{Zia05,Evans06}.
These models have in common a critical density $\rho^{crit}$ above which
a condensation transition occurs, that is, a finite fraction of the
total mass condenses onto a given site (or on a small domain
\cite{Evans06}). 

The standard way to compute the critical density in ZRP for instance,
is to use the grand-canonical partition function $\tilde{Z}$,
considered as a function
of the fugacity $z=e^{-\lambda}$ \cite{Evans-Rev05}:
\begin{equation}
\tilde{Z}(z)=\sum_{N=0}^{\infty}z^N Z_N(M)
\end{equation}
where $z$ is fixed by the density
\begin{equation}
\rho=\frac{M}{N}=\frac{z}{N}\frac{\partial \ln \tilde{Z}}{\partial z}\;,
\end{equation}
and to look for the convergence radius of $\tilde{Z}(z)$ in the complex plane
of $z$. To study the
condensed phase in more details, it is necessary to use a canonical ensemble
approach where the total mass is fixed \cite{Zia05}.
Accordingly, the present ITP formalism turns out to be well-suited
for such a study.

In Sec.~\ref{one_species} and \ref{two_species}, we illustrate in a
pedagogical manner how ITPs may give a
natural quantitative description of the condensation phenomenon, on the
example of simple mass transport models.
Qualitatively, the general procedure proposed is the following.
Interpreting the condensation as the coexistence of a fluid phase
and a condensed phase, one concludes that from the definition of ITPs,
the value of the ITPs should equalize in the two phases. The ITP for
the single-site condensate is often easily obtained, in which case the
value of the ITP in the fluid phase is also known.
Then the equation of state of the fluid phase, computed in the
grand-canonical ensemble, can be used to determine the density of this
phase (which turns out to be the critical density).
Hence, the total mass of the fluid phase is known.
>From the knowledge of the total mass, one finally deduces
the mass of the condensate.

Such a description of phase coexistence is a good illustration of
the application of ITPs to a rather simple out-of-equilibrium situation.
It is also a first step toward the description of the more complex
situation of the contact between two different systems. As we shall see
in Sec.~\ref{sec-contact}, this case may reveal some difficulties,
unexpected within a formal analogy with equilibrium (for instance
by defining a formal grand-canonical distribution, or through the introduction
of an effective hamiltonian), but well enlightened
in the present framework.

\subsection{Mass transport model with one species}
\label{one_species}

\subsubsection{Homogeneous model}

As a first example of the application of the ITP concept in phase coexistence
let us return to the model described in chapter \ref{fact}, defined by
the transport rates given in Eq.~(\ref{mtm-rate}). Assuming that
$f_i(m)=f(m)$ for all $i$, and that $f(m) \sim m^{-\gamma}$ for $m \to \infty$,
one finds for $\gamma>2$ a condensation if the average density
exceeds a critical value \cite{Evans-Rev05,Zia05}.
In the following we revisit this condensation transition, and show how our
approach of equalized ITPs can reveal the physics behind this phenomenon.
If we assume that the condensate,
which occurs on a randomly chosen site $j$, carries the macroscopic mass
$M_c = \mathcal{O}(N)$, its canonical partition function yields
\begin{eqnarray}
Z_c(M_c)&=&\int_0^{\infty} dm_j\, f(m_j)\, \delta(m_j-M_c)\nonumber\\
&=&f(M_c)\sim M_c^{-\gamma}\;.
\end{eqnarray}
The ITP conjugated to $M_c$ is the chemical potential $\lambda_c$
of the condensate, which is given by
\begin{equation}
\lambda_c = \frac{d \ln Z_c}{d M_c}
\approx -\frac{\gamma}{M_c}\;.
\end{equation}
In the thermodynamic limit $M_c \to \infty$, one thus obtains $\lambda_c=0$.
The equality of ITPs for the condensed and the fluid phase therefore leads
to $\lambda_f=\lambda_c=0$. The density $\rho_f$ of the fluid phase,
associated to
$\lambda_f$, can be determined from the equation of state computed in the
grand-canonical ensemble
\begin{equation} \label{rho-lambda-gc}
\rho_f = \frac{\int_0^{\infty}dm\,m f(m) e^{-\lambda_f m}}{\int_0^{\infty}dm\,f(m)e^{-\lambda_f m}}
=\frac{\int_0^{\infty}dm\,m f(m)}{\int_0^{\infty}dm\,f(m)}
\end{equation}
and exactly gives the critical density $\rho^{crit}$ \cite{Zia05}. Note that
for a value of $\gamma<2$ the critical density becomes infinite, which 
means that no condensation occurs.
Accordingly, the mass of the condensate is given by
\be
M_c = M - N\rho^{crit}\; ,
\ee
if the overall density $\rho=M/N$ is larger than $\rho^{crit}$.
This therefore leads to a thorough description of the condensation
in this system.

\subsubsection{Model with an impurity} \label{impurity-model}

Another well-known situation where condensation occurs is when a single
impurity site exhibits a dynamics which differs from those of the other sites
\cite{Evans-Rev05}. In this case the condensation no longer occurs on a
randomly chosen site, but on the impurity itself.
One of the simplest choice for the weights in
such a model corresponds to 
\begin{equation}
f_1(m) = f_{\rm imp}(m) = e^{\xi m}\quad \mbox{with}\quad \xi>0 
\end{equation}
for the impurity site and 
\begin{equation} \label{hom-phase}
f_i(m) = f_{\rm hom}(m) = m^{\eta-1}\quad \mbox{with}\quad \eta>0
\end{equation}
for the remaining sites $i>1$ ('hom' stands for 'homogeneous').
In this case the canonical partition function for the condensate reads
\begin{equation}
Z_c(M_c)=e^{\xi M_c}\;,
\end{equation}
which amounts to an ITP for the condensate 
\begin{equation}
\lambda_c=\frac{d \ln Z_c}{d M_c}=\xi.
\end{equation}
This yields for the fluid phase $\lambda_f=\xi$,
again by equalizing the ITPs of the two phases. 
The equation of state $\rho_f(\lambda_f)$
for the fluid phase, computed in the grand-canonical ensemble, reads
\begin{equation} \label{rho_f_imp}
\rho_f(\lambda_f) = \frac{\int_0^{\infty}dm\, m^{\eta} e^{-\lambda_f m}}{\int_0^{\infty}dm\, m^{\eta-1} e^{-\lambda_f m}}
= \frac{\Gamma(\eta+1)}{\lambda_f \Gamma(\eta)}
= \frac{\eta}{\lambda_f}\;.
\end{equation}
Thus the equality of the chemical potentials forces the
fluid to have a fixed density $\rho_0 = \eta/\xi$, as long as the condensate
is present, that is for $\rho=M/N > \rho_0$.
Interestingly, $\rho_0$ is not in itself the maximum density of the
homogeneous fluid phase, but is simply a density imposed by the impurity.

Finally, let us also briefly mention the interesting case where the
impurity is defined by
\be
f_{\rm imp}(m) = e^{\zeta m^2}
\ee
while the homogeneous phase is still defined by Eq.~(\ref{hom-phase}).
Assuming again that there is a condensate on the impurity, one finds
$\lambda_c = 2\zeta M_c \to \infty$. Thus one expects
$\lambda_f \to \infty$, yielding $\rho_f=0$ using Eq.~(\ref{rho_f_imp}).
This case is of particular interest, since on the one hand there is a
full localization of mass on the impurity in the thermodynamic limit,
and on the other hand
it cannot be studied within the framework of the grand-canonical ensemble,
due to the faster-than-exponential divergence of $f_1(m)$.

Accordingly, the above calculations provide
a simple description of the condensation in terms of the (out-of-equilibrium)
``equilibration'' of two coexisting phases. In particular, that the condensate
may absorb all the excess mass is understood due to the fact that $\lambda_c$
is independent of $M_c$.

\subsection{Mass transport model with two species}
\label{two_species}

The approach explained above for a model with one conserved quantity can easily be
extended to more complicated dynamics where more than one species are
involved.
In the case of two species, the general strategy is the following.
Let us consider a lattice model with two conserved masses
$M_1=\sum_{i=1}^N m_{1i}$ and $M_2= \sum_{i=1}^N m_{2i}$.
We assume that a condensate takes place on a given site $i_0$.
The key point of the approach is that the ITPs conjugated to $M_1$ and $M_2$
take equal values in both phases.
We shall denote these common values as $\lambda_k$, $k=1,2$, 
regardless of the phase we are dealing with.
Assuming a factorized steady-state, one can easily compute the values of $\lambda_k$
in the condensate as a function of the condensed masses $M_{1c}$ and $M_{2c}$,
leading to two equations of the form:
\be
\lambda_k = \lambda_{kc}(M_{1c},M_{2c})\;, \qquad k=1,2\; .
\ee
By analogy with the single-species case, it is expected that at least one
of the ITPs vanishes in the condensate in the thermodynamic limit, as long as
we consider a system without impurities; we thus assume that $\lambda_1 = 0$.
Then, one needs to compute the densities in the fluid phase using the
grand-canonical equation of state $\rho_{kf}(\lambda_1=0,\lambda_2)$.
Finally, the conservation of the total mass generically yields two coupled
non-linear equations for the unknown variables $M_{1c}$ and $M_{2c}$:
\be
M_{kc}+N\rho_{kf}(0, \lambda_{2c}(M_{1c},M_{2c})) = M_k\;, \quad k=1,2\; .
\ee
These equations may be hard to solve analytically in general, but they
could be solved numerically. In addition, simplifications
may appear in some cases, as we now illustrate on a specific model.

Let us consider a model with two conserved masses as above,
which are being transported on a lattice of arbitrary dimension $d$,
with periodic boundary conditions. The transfer rates for an amount of
mass $\mu_1$ or $\mu_2$ from a random site to a neighboring one are given
by $\vp_1(\mu_1|m_{1i},m_{2i})$ or $\vp_2(\mu_2|m_{1i},m_{2i})$
respectively, in accordance with 
\begin{eqnarray} 
\label{eq-phi1}
\varphi_1(\mu_1|m_{1i},m_{2i}) &=& v_1(\mu_1)\, \frac{h(m_{1i}-\mu_1,m_{2i})}{h(m_{1i},m_{2i})}\\
\label{eq-phi2}
\varphi_2(\mu_2|m_{1i},m_{2i}) &=& v_2(\mu_2)\, \frac{h(m_{1i},m_{2i}-\mu_2)}{h(m_{1i},m_{2i})}\\
\end{eqnarray}
For lattices of dimensions $d>1$, one allows a symmetric
mass transfer in any direction transverse to the flux with the same rates
$\vp_1(\mu_1|m_{1i},m_{2i})$ and $\vp_2(\mu_2|m_{1i},m_{2i})$ as in the
direction of the flux. The steady-state distribution in arbitrary dimension
$d$ is then given by
\begin{equation}
P(\{m_{ki}\}) = \frac{\prod_{i=1}^N g(m_{1i},m_{2i})}{Z(M_1,M_2)}
\prod_{k=1}^2 \delta \left( \sum_{i=1}^N m_{ki} - M_k \right)\;.
\end{equation}
Guided by the studies of
the two-species ZRP \cite{Hanney}, we choose an exponential form for the
weights 
\begin{equation}
h(m_1,m_2)=\exp(-\kappa m_1 m_2^{-\sigma})
\end{equation}
with $\kappa,\sigma>0$.
To obtain the conditions under which a condensation occurs, we calculate
the canonical partition function of the single-site condensate
$Z_c(M_{1c},M_{2c})=h(M_{1c},M_{2c})$. Hence one obtains for the
chemical potentials of the two species, computed in the condensate
\begin{eqnarray}
\lambda_1 = \frac{\partial \ln Z_c}{\partial M_{1c}}
&=& -\frac{\kappa}{M_{2c}^{\sigma}}\;,\\
\lambda_2 = \frac{\partial \ln Z_c}{\partial M_{2c}}
&=& \kappa\sigma\frac{M_{1c}}{M_{2c}^{\sigma+1}}\;.
\label{chem}
\end{eqnarray}
Therefore in the thermodynamic limit, one finds for the chemical potential
of the first species $\lambda_1=0$. The densities in the fluid phase
are then computed in the grand-canonical ensemble approach, which yields
for $\rho_{1f}$:
\begin{eqnarray}\label{dens2}
\rho_{1f}(0,\lambda_2) &=& \frac{\int_0^{\infty}dm_2\, e^{-\lambda_2 m_2}
\int_0^{\infty}dm_1\,
 m_1 e^{-\kappa m_1 m_2^{-\sigma}}}
{\int_0^{\infty}dm_2\, e^{-\lambda_2 m_2} \int_0^{\infty}dm_1\, 
e^{-\kappa m_1 m_2^{-\sigma}}}\nonumber\\
&=& \frac{1}{\kappa \lambda_2^{\sigma}}\frac{\Gamma(1+2 \sigma )}
{\Gamma(1+\sigma )}\;.
\end{eqnarray}
For $\rho_{2f}$, one obtains in the same way
\be
\rho_{2f}(0,\lambda_2) = \frac{\int_0^{\infty}dm_2 \, m_2 e^{-\lambda_2 m_2} \int_0^{\infty}
dm_1\,e^{-\kappa m_1 m_2^{-\sigma}}}{\int_0^{\infty}dm_2\, e^{-\lambda_2 m_2} 
\int_0^{\infty}dm_1\, e^{-\kappa m_1 m_2^{-\sigma}}}\nonumber\\
\ee
The integration over $m_1$ is straightforward and leads to
\begin{eqnarray}\label{dens1}
\rho_{2f}&=&\frac{1}{\lambda_2}\frac{\Gamma(2+\sigma)}{\Gamma(1+\sigma)}
= \frac{\sigma+1}{\lambda_2}\;.
\end{eqnarray}
Hence if a condensate exists, one has $\lambda_2>0$ to ensure a finite
positive density $\rho_{2f}$ in Eq.~(\ref{dens1}).
Then, from Eq.~(\ref{chem}) one finds that
$M_{2c}\sim M_{1c}^{1/(\sigma+1)}$, which means that the condensed mass
$M_{2c}$ behaves subextensively, and the density
of the fluid phase for the second species is thus equal to the overall
density, $\rho_{2f}=\rho_2$.
Due to Eq.~(\ref{dens1}) it follows that $\lambda_2$ is determined only
by the value of $\rho_2$. Therefore, the value of $\rho_{1f}$ is also fixed by
the overall density $\rho_2$ of the second species,
since it is only dependent on $\lambda_2$, as seen in Eq.~(\ref{dens2}).
As a result, a condensate forms only under the condition that
$\rho_1>\rho_1^{crit}(\rho_2)$ given by
\be
\rho_1^{crit}(\rho_2) = \rho_{1f}(0,\lambda_2(\rho_2))
= \frac{\Gamma(1+2\sigma)}{\kappa \, \Gamma(1+\sigma)} \,
\left(\frac{\sigma+1}{\rho_2}\right)^{\sigma}
\ee
Using Eqs.~(\ref{chem}) and (\ref{dens1}), as well as
the relation $M_{1c}=N(\rho_1-\rho_1^{crit})$, one finds the following
expression for the condensed mass of the second species
\begin{equation}
M_{2c} = \left[ \frac{\kappa\sigma \rho_2}{\sigma+1} N (\rho_1-\rho_1^{crit})
\right]^{\frac{1}{\sigma+1}}\;.
\end{equation}
To sum up, we have shown in this section that in the framework of ITPs,
a simple procedure
can be developed to describe the coexistence of different nonequilibrium
phases in mass transport models,
based on the idea that ITPs equalize their values
in these phases. We now turn to the more complex situation, where systems
with a different microscopic dynamics are put into contact.

\section{Contact of systems with different dynamics and ITP measure} \label{sec-contact}

\subsection{General approach} \label{sec-cont-genap}

Within the equilibrium context, the temperature, pressure and chemical
potential of two different systems put into contact equalize, as long
as the contact allows the conjugated conserved quantity (energy, volume
or particles) to be exchanged. This is actually a strong statement,
as it is true even if the two systems considered have very different
microscopic dynamics, provided that both systems can be described by an
hamiltonian.
Whether such property also holds in the present more general context
of nonequilibrium ITP formalism is thus an essential issue. Potentially
interesting applications are the description of the effect of the environment
on a system, or the possibility to measure the value of an ITP using a
small auxiliary system, in the same way as temperature is measured with
a thermometer.

As we shall see in this section, the equalization of the values of the ITPs
in two systems in contact is actually not automatically fulfilled
in a nonequilibrium context due to the necessity to satisfy the additivity
condition (\ref{afp}), (\ref{afp2}), and the way the two systems are connected
has to be considered carefully.
To highlight this point,
let us examine two different systems (that is, with different microscopic
dynamics) $\mathcal{S}_1$ and $\mathcal{S}_2$ that
separately conserve the same physical quantities $Q$, with values $Q_1$ and
$Q_2$ respectively. 
When put into contact,
the dynamics at the interface generates the distribution $\Phi(Q_1|Q)$
for the random partition of $Q=\sum_i q_i$ into $Q_1$
(in system $\mathcal{S}_1$) and $Q_2=Q-Q_1$ (in system
$\mathcal{S}_2$) respectively.
Assuming, in the spirit of equilibrium calculations, that the two systems
are weakly coupled, i.e.~the only coupling is the exchange of $Q$,
the global probability distribution reads
\bea \nonumber
P(\{q_i\}) &=& \int_0^Q dQ_1\, \Phi(Q_1|Q) P_1(\{q_{i_1}\}) \,
P_2(\{q_{i_2}\})\\ \nonumber
&=&\frac{\Phi(\sum_{i_1 \in \mathcal{S}_1} q_{i_1}|Q)}
{Z_1(\sum_{i_1 \in \mathcal{S}_1} q_{i_1})
Z_2(Q-\sum_{i_1 \in \mathcal{S}_1} q_{i_1})}\\
&\times& F_1(\{q_{i_1}\})\, F_2(\{q_{i_2}\}) \, \delta \left(\sum_i q_i-Q \right)
\label{Pqi-contact}
\eea
where $F_1(\{q_{i_1}\})$ and $F_2(\{q_{i_2}\})$ are the probability weights
of systems $\mathcal{S}_1$ and $\mathcal{S}_2$ respectively, taken as
isolated.
For the ITP to equalize, the additivity condition
(\ref{afp}), (\ref{afp2}) has to be satisfied, when applied to $\mathcal{S}_1$ and
$\mathcal{S}_2$ considered as the two subsystems of the global system.
Intuitively, this additivity condition means that the probability weight
essentially factorizes, as already mentioned when describing the
grand-canonical ensemble.
Due to the appearance of the quantity $\sum_{i_1 \in \mathcal{S}_1} q_{i_1}$
in Eq.~(\ref{Pqi-contact}), nonlocal contributions arise in the
probability weights appearing in $P(\{q_i\})$.
For these weights to factorize (and thus for the additivity condition
to be fulfilled), the prefactor
$\Phi/Z_1 Z_2$ has to be equal to a constant (since it would otherwise
depend on the nonlocal quantity $\sum_{i_1 \in \mathcal{S}_1} q_{i_1}$)
which is nothing
but $1/Z(Q)$, up to corrections vanishing in the thermodynamic limit.
Thus $\Phi(Q_1|Q)$ should necessarily be of the form
\begin{equation} \label{equaliz-cdtn}
\Phi(Q_1|Q) \approx \frac{Z_1(Q_1)Z_2(Q-Q_1)}{Z(Q)}\;,
\end{equation}
again up to possible small corrections.
This result may alternatively be interpreted in the following way.
The distribution $\Phi(Q_1|Q)$ is actually nothing but the conditional
distribution $\Psi(Q_1|Q)$ introduced in Eq.~(\ref{afp}), for the
specific partition of the global system into $(\mathcal{S}_1, \mathcal{S}_2)$
--whereas $\Psi(Q_1|Q)$ is a priori defined for an arbitrary partition.
Eq.~(\ref{equaliz-cdtn}) is then simply the additivity condition
(\ref{afp}), (\ref{afp2}) applied to the partition $(\mathcal{S}_1, \mathcal{S}_2)$.

When putting into contact two nonequilibrium systems,
$\Phi(Q_1|Q)$ does not obey Eq.~(\ref{equaliz-cdtn}) in general,
so that ITPs do not necessarily equalize.
Yet for special cases, this equalization may be
recovered, as we shall see in Sec.~\ref{sect-connect-MTM}.
In any case, it is an important challenge to be able to understand which
microscopic properties the dynamics of the contact has to satisfy so that
equalization of ITPs hold. This is the topic of Sec.~\ref{carac-contact}.

To clarify the relation with the equilibrium case, let us
consider for instance the equilibrium canonical ensemble,
where the conserved quantity would be the number of particles.
The system is described by a Boltzmann-Gibbs distribution, and assuming
that the hamiltonian does not include long-range interactions,
the additivity condition (\ref{afp}), (\ref{afp2}) is satisfied.
>From this, one deduces that Eq.~(\ref{equaliz-cdtn}) is necessarily
satisfied, since otherwise the additivity condition would not hold.

\subsection{Connecting two different mass transport models}
\label{sect-connect-MTM}

\begin{figure}[t]
\centering\includegraphics[width=7cm,clip]{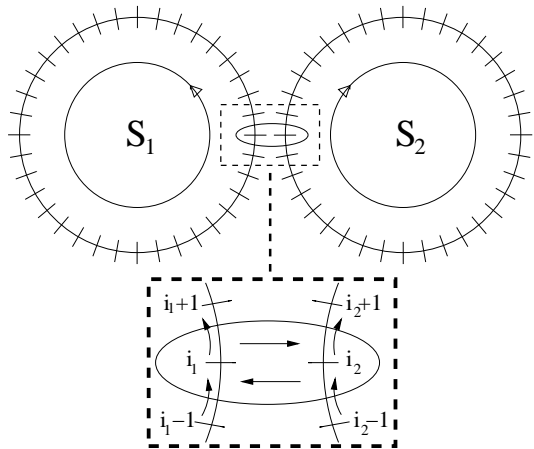}
\caption{Illustration of two mass transport models with periodic boundary
conditions, connected through a contact made of a single additional link.
A zoom over the contact is also shown in the dashed frame; arrows indicate
the oriented links through which mass is transported.
}
\label{fig-contact}
\end{figure}

To illustrate an implementation of the above idea, we consider two
single-species
mass transport models in contact, as represented schematically
in Fig.~\ref{fig-contact}. The dynamical rules are
site-independent within each system, but are different in 
$\mathcal{S}_1$ and $\mathcal{S}_2$. To be more specific,
the transport rate within system $\mathcal{S}_{\nu}$ reads
\be \label{MTM-contact}
\varphi_{\nu}(\mu|m) = v(\mu)\frac{f_{\nu}(m-\mu)}{f_{\nu}(m)}\;,\qquad
\nu=1,2\; .
\ee
Let us emphasize that the function $v(\mu)$ is the same for both systems.
The reason for this choice, ensuring the equalization of the ITPs, will
be explained in detail in Sec.~\ref{carac-contact}.
A particular case where this condition holds is the ZRP one, where $\mu=1$
is the only allowed value, and $v(1)$ is set to $1$.

In order to define the dynamics at the contact,
we extend the dynamical rules in such a way that a mass
located at site $i_1 \in \mathcal{S}_1$ at the interface (see Fig.~\ref{fig-contact}), 
is transferred with rate $\varphi_1(\mu|m)$
either to the neighboring site $i_1+1$ or with the same rate to site
$i_2\in \mathcal{S}_2$. A similar rule is applied 
for a mass located on site $i_2$, regarding the transfer to $i_1$.
Hence the two
homogeneous systems simply combine to an inhomogeneous one, for which one
can find a factorized steady-state, as $v(\mu)$ is
site-independent, and as the graph on which the system is defined
satisfies the required geometrical constraint
\cite{graph-MTM}.
It follows that in this case the additivity condition (\ref{afp}), (\ref{afp2}) holds,
and the ITPs of the two systems equalize.

Let us now give some examples illustrating the consequences
of the equalization of ITPs for two systems in contact, within the framework
of the above model.
It should first be noticed that the equalization of ITPs enforces
constraints on the densities of the two systems,
which may thus differ one from the other if the two systems have different
equations of state.
An interesting situation arises in the presence of a condensate.
To be more specific, let us consider two different mass transport models,
without impurities. The two systems are initially both
isolated, and we assume that the first one contains a condensate, while the
second one does not. This means that, in this initial stage, $\lambda_1=0$
whereas $\lambda_2>0$.
When put into contact, the dynamics will be such as to equalize the ITPs
$\lambda_1$ and $\lambda_2$, through a transfer of mass between the two
systems. As $\lambda$ generically decreases when the density increases,
mass has to be transferred from $\mathcal{S}_1$ to $\mathcal{S}_2$.
Withdrawing mass from $\mathcal{S}_1$ actually does not affect the fluid
phase of $\mathcal{S}_1$ (at least in a quasi-static limit)
as long as the average density $\rho_1$ remains
above the critical density $\rho_1^{crit}$, and mass is taken from the
condensate only, in this first stage.
Assuming that the critical density of $\mathcal{S}_2$, $\rho_2^{crit}$,
is infinite, the condensate in $\mathcal{S}_1$
eventually disappears, and both ITPs converge to a strictly positive
value that lies between the initial values of $\lambda_1$ and $\lambda_2$.
Yet, the final densities of the two systems are different in general.

\begin{figure}[t]
\centering\includegraphics[width=7.5cm,clip]{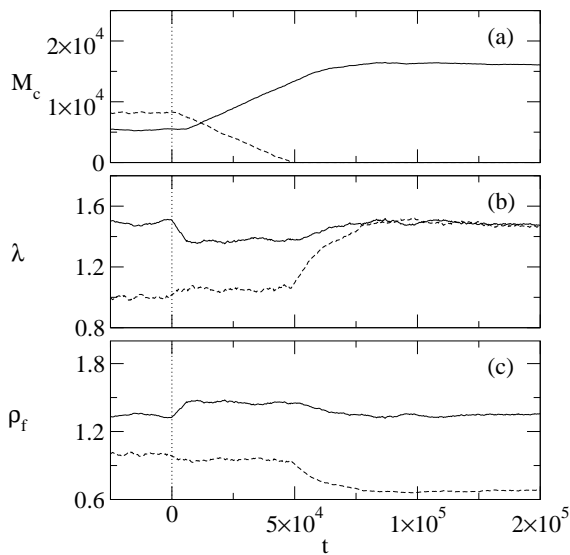}
\caption{Time dependence of the masses of the condensates (a), the ITPs (b)
and the densities of the fluid phases (c)
in two mass transport models in contact, $\mathcal{S}_1$
(solid line) and $\mathcal{S}_2$ (dashed line) --see
Fig.~\ref{fig-contact}. Both of them contain an impurity site.
Parameter values are $\xi_1=1.5$ and $\xi_2=1$ for the impurities,
$\eta_1=2$ and $\eta_2=1$ for the homogeneous parts --see text.
The systems are separately in steady state (with overall initial densities
$\rho_1=\rho_2=2$) before the contact is
established at time t=0 (dotted line).
Note that the ITPs eventually equalize, and that the difference
between the fluid densities increases after the contact is established.}
\label{fig-connected}
\end{figure}

One may also think of more complex situations. Let us consider the case
where each of the two systems contains an impurity
(as defined in Sec.~\ref{impurity-model}), with respective parameters
$\xi_1$ and $\xi_2$ ($\xi_2<\xi_1$). In addition, we consider
different functions $f_{\rm hom}(m)$ for the homogeneous part of
each system, namely $f_{{\rm hom},\nu}(m)=m^{\eta_{\nu}-1}$
in system $\mathcal{S}_{\nu}$, $\nu=1,2$.
Before the contact is switched on, both systems are in steady state and
contain a condensate, so that $\lambda_1=\xi_1$ and $\lambda_2=\xi_2$.
When the contact is established at $t=0$, the ITPs tend to equalize, and as
$\lambda_2 < \lambda_1$, mass is transferred from $\mathcal{S}_2$ to
$\mathcal{S}_1$ (as $\rho$ decreases with $\lambda$),
independently of the values of $\eta_1$ and $\eta_2$.
But as $\mathcal{S}_1$ already
contains a condensate, the density of the fluid phase is fixed, and 
all the mass brought to $\mathcal{S}_1$ is actually
transferred to the condensate. Hence, the final state corresponds to
$\lambda_1=\lambda_2=\xi_1$, with $\mathcal{S}_1$ containing a larger
condensate than initially, and $\mathcal{S}_2$ being at a density
$\rho_2(\lambda_2=\xi_1)<\rho_2^{crit}=\rho_2(\lambda_2=\xi_2)$,
so that there is no more condensate in $\mathcal{S}_2$.
Fig.~\ref{fig-connected} presents the results of numerical simulations
of the above situation, showing in particular that the condensate of
$\mathcal{S}_2$ disappears once the contact is established.
In addition, this simulation confirms that the ITPs of the two systems
are controlling the direction of the flux of mass. Indeed, mass
is transferred from $\mathcal{S}_2$ to $\mathcal{S}_1$ even though the
density of the fluid is larger in $\mathcal{S}_1$ than in $\mathcal{S}_2$,
which might seem rather counterintuitive given that the two systems
are in contact through their fluid phases only.

Note also that strictly speaking, the determination of $\lambda$
using the equation of state is valid only in a steady state, so that
one should wait until the density $\rho$ is stationary
before determining $\lambda$ from $\rho$.
Still, for the purpose of illustration, we present here $\lambda(t)$
deduced from the fluid density $\rho_f(t)$ even in the nonstationary regime.
This is meaningful if the exchanges between the two systems are slow enough
so that each system may be considered in a quasi-steady state.

>From the above result, one sees that the impurity in system
$\mathcal{S}_1$ (that is, the one with the larger value of $\xi$)
plays the role of a reservoir of mass that fixes the value of the ITP
of $\mathcal{S}_2$ to $\lambda_2=\xi_2$.
Interestingly, $\mathcal{S}_1$ only needs to contain
a mass of the same order as that of $\mathcal{S}_2$ to act as
a reservoir, as the ITP $\lambda_{1c}$ of the condensate
is independent of its mass $M_{1c}$ as long as $M_{1c}$ is
macroscopic.
On the contrary, usual reservoirs require a mass much larger than
that of the systems they are in contact with.

As we have seen on these simple examples, the notion of ITP allows one
to make, essentially without calculations, non-trivial predictions about
the behavior of mass transport models put into contact
(for instance, the final densities of both systems); only the knowledge
of the equation of state for each system taken separately is required.
However, such predictions can be made only if the ITPs equalize.
In the present case of mass transport models, this equalization holds
as long as the transition rates obey the relation (\ref{MTM-contact}).
What are the conditions for such an equalization to hold in more general
situations will be the topic of the next section.

\subsection{Characterization of the dynamics at the contact}
\label{carac-contact}

\begin{figure}[t]
\centering\includegraphics[width=7cm,clip]{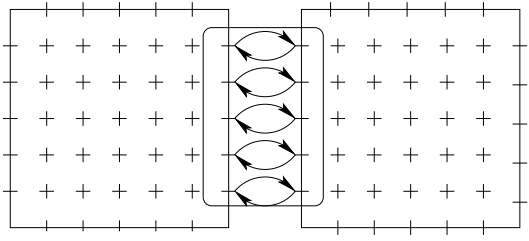}
\caption{Schematic drawing of two generic systems in contact.
Arrows illustrate the possibility to transfer a conserved quantity
between the two systems.
Contrary to Fig.~\protect\ref{fig-contact}, the contact
generally extends over many sites.}
\label{fig-contact-complex}
\end{figure}

In this section, we consider the more general case of two systems in
contact as schematically illustrated in Fig.~\ref{fig-contact-complex}.
The dimension $d$ of the two systems is arbitrary, and the contact consists
in a set of links between the two systems. Generically, the contact relates
some parts of the borders of each system, as shown in
Fig.~\ref{fig-contact-complex},
but one may also think of more complex types of contact.

As mentioned in Sec.~\ref{sec-cont-genap}, the dynamics at the contact
plays an essential role in the possibility to equalize the ITPs
of two connected systems. This equalization occurs if the condition
(\ref{equaliz-cdtn}), or equivalently, the additivity condition
(\ref{afp}), (\ref{afp2}), is fulfilled. Yet, testing this condition is a priori
very difficult, and its interpretation in terms of the microscopic
dynamics at the contact is not obvious.
On the other hand, if two systems $\mathcal{S}_1$ and $\mathcal{S}_2$
are connected and globally isolated, the flux $\phi_{1\to 2}$ transferred
from $\mathcal{S}_1$ to $\mathcal{S}_2$ through the contact has to be equal,
in steady state, to the reverse flux $\phi_{2\to 1}$ going from
$\mathcal{S}_2$ to $\mathcal{S}_1$.

For the sake of simplicity, we now introduce some assumptions
that allow one to analyze, at least in a simple case,
this generic problem of the contact between two systems.
For definiteness, let us consider lattice systems $\mathcal{S}_1$
and $\mathcal{S}_2$ that may exchange a globally conserved quantity.
The set of sites belonging to the contact
in system $\mathcal{S}_{\nu}$, $\nu=1,2$, is denoted as $\mathcal{C}_{\nu}$.
As already mentioned, the situations considered correspond to the weak
coupling limit. In the present context, this means that
the flux per site crossing the contact is typically much smaller than
the local flux between two sites of a given system, so that the contact
does not perturb the dynamics of each system, apart from the (slow) exchange
of $Q$. In addition,
the specific assumptions used in the following arguments are that:

(i) the flux $\phi_{1\to 2}$
depends only on $\lambda_1$, and not on the properties of $\mathcal{S}_2$
such as $\lambda_2$; respectively, $\phi_{2\to 1}$ depends only on
$\lambda_2$ (yet, the total flux $\phi_{1\to 2}-\phi_{2\to 1}$ depends
on both $\lambda_1$ and $\lambda_2$);

(ii) the probability weights of $\mathcal{S}_1$ and
$\mathcal{S}_2$, each one considered as isolated, are factorized as products
of one-site weights.

In the spirit of hypothesis (i), 
the dynamics at the contact is defined by the probability rate
$\varphi_{i_1}^c(\mu|q_{i_1})$ to transfer $\mu$ from
site $i_1 \in \mathcal{C}_1$ in $\mathcal{S}_1$ to $\mathcal{S}_2$;
a similar rate $\varphi_{i_2}^c(\mu|q_{i_2})$ defines the transfer
from $i_2 \in \mathcal{C}_2$ in
$\mathcal{S}_2$ to $\mathcal{S}_1$.
Under these assumptions, one can compute the fluxes $\phi_{1\to 2}(\lambda_1)$
and $\phi_{2\to 1}(\lambda_2)$;
in particular, $\phi_{1\to 2}(\lambda_1)$ reads
\be
\phi_{1\to 2}(\lambda_1) = \sum_{i \in \mathcal{C}_1} \int_0^{\infty}
dq_i\, P_1(q_i) \int_0^{q_i} d\mu\,\mu\, \varphi_i^c(\mu|q_i).
\ee
where $P_1(q_i)$ is the single site probability distribution in
$\mathcal{S}_1$. Given that $P_1(q)=c_1 f_1(q) \exp(-\lambda_1 q)$,
$c_1$ being a normalization constant, it follows that
\be \label{flux-lambda}
\phi_{1\to 2}(\lambda_1) = c_1 \int_0^{\infty} dq
\int_0^q d\mu \, \mu \varphi_1^{tot}(\mu|q) f_1(q)\, e^{-\lambda_1 q}\;,
\ee
where we have introduced the ``total'' rate
\be
\varphi_1^{tot}(\mu|q) = \sum_{i \in \mathcal{C}_1} \varphi_i^c(\mu|q)\;.
\ee
By exchanging the indexes of the systems, a similar relation holds for
$\phi_{2\to 1}(\lambda_2)$.

For the two systems to equalize their ITPs, it is necessary that the
equality $\phi_{1\to 2}(\lambda_1) = \phi_{2\to 1}(\lambda_2)$ leads to
$\lambda_1=\lambda_2$, for arbitrary values of (say) $\lambda_1$.
In other words, the two functions
$\phi_{1\to 2}(\lambda)$ and $\phi_{2\to 1}(\lambda)$ must be identical.
Let us then compute the difference $\phi_{1\to 2}(\lambda)-
\phi_{2\to 1}(\lambda)$. After a straightforward calculation, this difference
can be expressed as
\bea \nonumber
&&\phi_{1\to 2}(\lambda)-\phi_{2\to 1}(\lambda) =
c_1 c_2 \int_0^{\infty} dq_1 \int_0^{\infty} dq_2 \int_0^{q_1} d\mu \\
\nonumber
&\times& \mu \,  e^{-\lambda (q_1+q_2)}
[ \varphi_1^{tot}(\mu|q_1) f_1(q_1) f_2(q_2) \\
&& \qquad - \varphi_2^{tot}(\mu|q_2+\mu) f_1(q_1-\mu) f_2(q_2+\mu)]
\eea
In order that $\phi_{1\to 2}(\lambda)-\phi_{2\to 1}(\lambda)$ vanishes for
any value of $\lambda$, it is necessary and sufficient that the expression
between brackets vanishes, that is
\bea \label{effect-DB}
&&\varphi_1^{tot}(\mu|q_1) f_1(q_1) f_2(q_2)\\ \nonumber
&& \qquad \qquad = \varphi_2^{tot}(\mu|q_2+\mu) f_1(q_1-\mu) f_2(q_2+\mu)
\eea
One can check that Eq.~(\ref{effect-DB}) is precisely a detailed
balance relation between configurations
$(q_1,q_2)$ and $(q_1'=q_1-\mu,q_2'=q_2+\mu)$. Yet, let us emphasize
that this detailed balance relation does {\it not} concern the
true {\it microscopic} dynamics at the contact, but rather an {\it effective,
coarse-grained} dynamics, defined by $\varphi_{\nu}^{tot}(\mu|q)$
that reduces the contact to a single effective link.
To illustrate this point, one can imagine a contact made of two fully
biased links between $\mathcal{S}_1$ and $\mathcal{S}_2$. The first link
can only transfer the conserved quantity $Q$ from $\mathcal{S}_1$ to
$\mathcal{S}_2$, whereas the second one only allows $Q$ to be transferred
from $\mathcal{S}_2$ to $\mathcal{S}_1$. Then, the microscopic
dynamics at the contact does not satisfy detailed balance, but the
coarse-grained dynamics may fulfill this condition.

Note that if the true contact already consists in a single link
as illustrated on Fig.~\ref{fig-contact}, the effective dynamics is
the true one,
so that the dynamics at the contact really satisfies detailed balance.
This is basically the interpretation of the condition used in
Eq.~(\ref{MTM-contact}), according to which $v(\mu)$ has to be the same
in the two systems:
this condition ensures detailed balance at the contact, even though
detailed balance breaks down in each system due to the presence of fluxes.
In constrast, if the functions $v(\mu)$ were to be different in the
two systems the ITPs would not equalize due to the lack of detailed balance
at the contact.

In summary, we have seen the importance of the dynamics of the contact,
which must fulfill condition (\ref{equaliz-cdtn}) to ensure the
equalization of the ITPs among the connected systems.
On the other hand, the effective detailed balance Eq.~(\ref{effect-DB})
is a priori a less general statement, due to the simplifying assumptions
used to derive it.
However it provides us with a physical interpretation of the conditions
required for the dynamics at the contact. One may hope that the resulting
physical picture might be relevant beyond the strict validity of
Eq.~(\ref{effect-DB}).

\subsection{Measure of an ITP}

\subsubsection{Notion of ITP-meter}
\label{notion}

An essential issue about ITPs would be the ability to measure them.
In analogy to the measurement of temperature in equilibrium statistical
mechanics, one could imagine to realize such a measurement by connecting
an auxiliary system to the system under consideration. This problem has
also been addressed in \cite{Hatano}.
In this general framework, we shall call this (perhaps conceptual)
instrument ``ITP-meter'', in reference to the nomenclature ``thermometer''.
There are three essential requirements which have
to be fulfilled to perform such an implementation.
First, the values of $\lambda$ in the gauged system and in the ITP-meter
need to equalize. Second, the measured system
must not be disturbed by the measurement, so that the ITP-meter should be
small with respect to the system upon which the measure is performed.
Third, the equation of state of the ITP-meter has to be known, since 
one needs to deduce the value of its ITP from the measure of a directly
accessible physical quantity.
Accordingly, these different conditions turn the realization of an
ITP-meter into a highly nontrivial problem.

\begin{figure}[t]
\centering\includegraphics[width=7.5cm,clip]{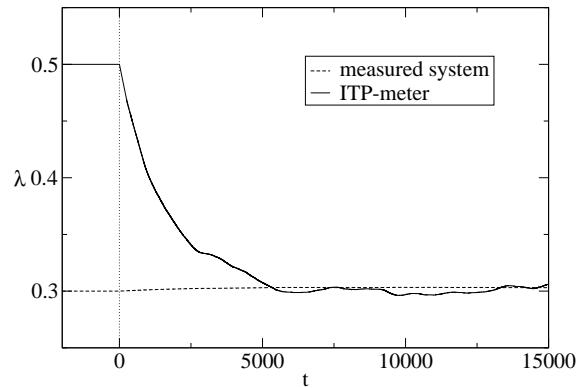}
\caption{Measurement process on a system $\mathcal{S}_1$ using an ITP-meter
$\mathcal{S}_2$. The values of the ITPs $\lambda_1$ (dashed line) and
$\lambda_2$ (solid line) are plotted versus time.
The systems are characterized by parameters
$\eta_1=3$, $\eta_2=5$ (see text). The contact is switched on at time
$t=0$.
Initially, $\rho_1=\rho_2=10$, so that $\lambda_1=0.3$ and $\lambda_2=0.5$.
Note that $\lambda_1$
almost stays constant whereas after a sufficient time $\lambda_2$ 
converges to $\lambda_1$.}
\label{fig-itp-meter}
\end{figure}

Nevertheless it is possible for certain simple cases to realize a measurement
with an ITP-meter, using for instance two connected mass transport models
as described in Sec.~\ref{sect-connect-MTM}.
The systems are similar to the ones shown in Fig.~\ref{fig-contact},
but now $\mathcal{S}_2$, used as an ITP-meter,
is much smaller than $\mathcal{S}_1$.
We present in Fig.~\ref{fig-itp-meter} numerical simulations in which
each system $\mathcal{S}_{\nu}$ ($\nu=1,2$) is homogeneous, and obeys the
transport rate defined in Eq.~(\ref{MTM-contact}) with
$f_{\nu}(m)=m^{\eta_{\nu}-1}$, where $\eta_1 \ne \eta_2$.
By measuring the density $\rho_{\nu}$ of each system, one can determine
the value of $\lambda_{\nu}$ using the equations of state,
namely $\lambda_{\nu}=\eta_{\nu}/\rho_{\nu}$.
In practice, one would of course only measure the density of the ITP-meter,
but here we determine both $\lambda_1$ and $\lambda_2$ to check the validity
of the approach.

Numerical simulations are run using systems of size $N_1=64510$ and
$N_2=1024$, with equal initial densities,
$\rho_1=\rho_2=\rho$, and with $\eta_1 < \eta_2$.
Hence the initial values of the ITPs are different,
namely $\lambda_1=\eta_1/\rho$ and $\lambda_2=\eta_2/\rho$.
At time zero the contact between the two systems is switched on,
and mass flows from $\mathcal{S}_1$ to the ITP-meter $\mathcal{S}_2$,
since $\lambda_1<\lambda_2$.
As predicted theoretically and confirmed by the numerical simulations,
the ITPs of the two systems equalize once the steady-state is reached.
In addition, the value of $\lambda_1$ does not change significantly along
this process, which is the basic requirement for a non-perturbative
measurement. Accordingly, the ITP-meter indeed measures the value of
$\lambda_1$. Quite importantly, this measure is done without knowing
the value of the parameter $\eta_1$ defining the dynamics of
$\mathcal{S}_1$. This value was only used to determine $\lambda_1$ from
$\rho_1$ in order to check the measurement.

The above example illustrates on a simple model that it is in principle
possible to measure an ITP using an ITP-meter in nonequilibrium systems.
Still, in more realistic situations, finding a suitable definition for the
dynamics at the contact that allows for the equalization of ITPs
turns out to be a major challenge. As seen in the preceding section,
Eq.~(\ref{effect-DB}) gives a condition for the equalization
of ITPs between the two systems, and it provides useful information to
design the contact.
Yet, it is important to notice that Eq.~(\ref{effect-DB}) is a
detailed balance relation with respect to the stationary distributions of
$\mathcal{S}_1$ and $\mathcal{S}_2$.
Hence, to satisfy this relation,
one requires some important information on the gauged system,
namely its steady state weight $f_1(q)$. Such an information is usually
unavailable in nonequilibrium systems, contrary to what happens in
equilibrium, where the weights are either uniform
(microcanonical ensemble) or given by the Boltzmann-Gibbs factor
(canonical ensemble).

\subsubsection{Measure within subsystems}
\label{sub-measure}

An alternative route, that has been exploited recently in the context of
granular matter \cite{Nowak,Swinney}, consists in trying to determine
the ITP through
an interpretation of direct measurements on the system, instead of using
an auxiliary system (the ITP-meter) \footnote{Note that, however, the
procedure presented here is slightly different from that used in
\cite{Nowak,Swinney}.}.
More precisely, let us consider a system with a globally conserved quantity
$Q$. Then, from Eq.~(\ref{cum}), the variance of the quantity $Q_N$, measured
over a ``mesoscopic'' subsystem of size $N$, obeys the following relation,
derived from the grand-canonical ensemble:
\begin{equation}\label{var}
\langle Q_N^2 \rangle - \langle Q_N \rangle^2 = -\frac{d \langle Q_N \rangle}{d\lambda}\;.
\end{equation}
Let us assume, consistently with the additivity condition
(\ref{afp}), (\ref{afp2}), that the variance of $Q_N$ is linear in the subsystem
size $N$, at least for $1 \ll N \ll N_{\rm tot}$, where $N_{\rm tot}$
is the size of the global system.
Hence the variance may be written as
\begin{equation}
\langle Q_N^2 \rangle - \langle Q_N \rangle^2 = N g(\rho)
\end{equation}
with $\rho=\langle Q_N \rangle/N$. Then Eq.~(\ref{var}) leads to
\begin{equation}
g(\rho)=-\frac{d\rho}{d\lambda}\;.
\label{g_rho}
\end{equation}
Given a reference point $(\rho_0,\lambda_0)$, one can then determine the
equation of state of the system, simply by integrating Eq.~(\ref{g_rho})
numerically:
\begin{equation}
\lambda = \int_{\rho}^{\rho_0} \frac{d\rho}{g(\rho)}
+ \lambda_0\;.
\end{equation}

\begin{figure}[t]
\centering\includegraphics[width=7.5cm,clip]{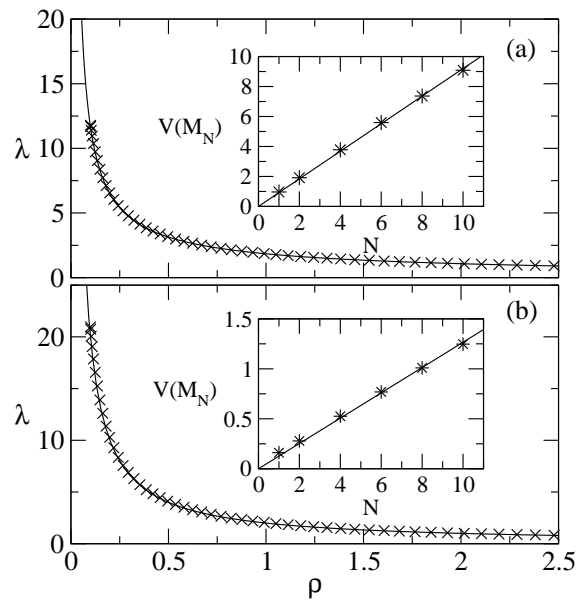}
\caption{Equation of state $\lambda$ as a function of $\rho$ in two mass
transport models, obtained by numerical simulations using the measurement
procedure described in the text ($\times$), compared with the
theoretical prediction $\rho(\lambda)$ derived in the grand-canonical
ensemble (solid line);
(a) fully factorized steady state with $f(m)=(1+m)^2$,
(b) pair-factorized case with $g(m,n)=m+n$.
Insets: dependence of the variance $V(M_N)$ on the
subsystem size $N$ ($\ast$) for densities $\rho=1.2$ (a) and $\rho=0.5$ (b);
solid lines are linear fits.
}
\label{fig-lambda-rho}
\end{figure}

As a first example, we now apply this procedure 
to a mass transport model on a ring
with dynamics defined by $f(m)=(1+m)^2$ and $v(\mu)=1$
(see Eq.~(\ref{mtm-rate})),
for which the equation of state $\lambda(\rho)$ cannot be determined
easily by scaling arguments as in Sec.~\ref{homo-model}.
By measuring in a numerical simulation the variance of the mass $M_N$ over
subsystems of different size $N$,
we find as expected a linear behavior in $N$ already for small sizes
(see inset of Fig.~\ref{fig-lambda-rho}(a)),
since the grand-canonical distribution is fully factorized.
The result of the numerical integration for the equation of state 
is in very good agreement with the theoretical curve obtained using the
grand-canonical ensemble (see Fig.~\ref{fig-lambda-rho}(a)), for which $\rho$
can be determined as a function of $\lambda$, as done for instance
in Eq.~(\ref{rho-lambda-gc}).

The fact that this procedure works as well for
cases in which the steady state does not fully factorize, can be
seen in another numerical experiment exhibiting a pair-factorized
steady state. Let the dynamics be defined as in Eq.~(\ref{pair-rate}),
with $v(\mu)=1$ and $g(m,n)=m+n$, that is, we assume $\alpha=1$,
$\beta=0$ and $\gamma=1$ in Eq.~(\ref{g-abc}).
Measuring again the variance of the mass
$M_N$ over subsystems of size $N$, shows that the measurement of the
ITP $\lambda$ should be performed for higher values of $N$ than in the
fully factorized case to have access to the linear regime (see inset of
Fig.~\ref{fig-lambda-rho}(b)).
Note that the results of the measurement (see Fig.~\ref{fig-lambda-rho}(b)) 
agree equally well with the theoretical curve, which can be obtained
in this case by the simple scaling argument presented in Sec.~\ref{pair}.

In these examples, we obtained for simplicity a reference point
($\rho_0, \lambda_0$) using the theoretical equation of state.
In a more realistic situation where the equation of state is unknown,
one can estimate as well a reference
point only through the information obtained from the measured variances.
Measuring numerically the function $g(\rho)$ for large values of $\rho$,
one can fit its asymptotic (large $\rho$) expression with a power law,
$g(\rho) \approx A \rho^{\delta}$; in the two models above, one has $\delta=2$.
Then, assuming that $\lambda$ vanishes when $\rho \to \infty$, and
that $\delta>1$, the reference value $\lambda_0$ corresponding
to a given large density $\rho_0$ is obtained as
\be
\lambda_0 = \int_{\rho_0}^{\infty} \frac{d\rho}{g(\rho)}
\approx \int_{\rho_0}^{\infty} \frac{d\rho}{A \rho^{\delta}}
= \frac{1}{A(\delta-1)\rho_0^{\delta-1}} \; .
\ee
Note that alternatively, one may also determine a reference point from the
distribution of $Q_N$, as proposed in \cite{Dauchot}.

Accordingly, the present approach provides a rather simple way to measure
experimentally or numerically
ITPs in realistic systems for which no information on the microscopic
probability distribution (like the weight factors $f(m)$ for instance)
is available. As mentioned above, this method has already been used in
the context of granular material \cite{Nowak,Swinney}. However, it was
thought to rely on Edwards' thermodynamic construction which explicitely
assumes the equiprobability of states compatible with the constraints
(conserved quantities), whereas this assumption is not necessary,
as illustrated in the above examples --see also \cite{Dauchot}.
Indeed, the validity of the
approach is much more general, making it a convenient way to
determine practically ITPs in nonequilibrium systems.

\section{Conclusions}
\label{sec-conclusion}

In this paper, we highlight that
the notion of generalized ITPs for steady-state nonequilibrium systems
serves as a new tool in the study of nonequilibrium phenomena.
Within the range of validity of the additivity condition
(\ref{afp}), (\ref{afp2}),
there is a vast variety of models for which our approach can be useful, like 
models with a factorized steady-state distribution, or one-dimensional
systems described by a matrix product ansatz or the transfer matrix method
(at least with finite matrices).
The method should also be of interest for models
which are well approximated by a mean field approach.

Introduced by generalizing the equilibrium concept, for which 
physical intuition is well developed, ITPs benefit
from a rather clear physical
interpretation.
They provide a convenient way to describe the coexistence of two phases,
as may be easily illustrated on the case of the condensation transition.
An essential issue, which made the success of the ITP concept in equilibrium,
is the possibility to equalize ITPs in different systems put into contact.
Contrary to equilibrium situations, such an equalization
is not automatically fulfilled in the nonequilibrium case, and the
dynamics at the contact turns out to play a major role.
We have derived, under simplifying assumptions, a coarse-grained detailed
balance relation that needs to be satisfied by the contact dynamics
in order that ITPs equalize.
Deriving a more general criterion by relaxing some of the
assumptions made would be interesting to gain further insights on this
important issue.

A related difficulty with the approach is that the additivity property
(\ref{afp}), (\ref{afp2}) may be hard to test directly. Still, measuring the variance
of the globally conserved quantity over subsystems of a homogeneous
system may lead to an indirect test of this property:
indeed, one expects that the variance introduced in Eq.~(\ref{var}), becomes linear in $N$ for $N \gg 1$ if the additivity property is satisfied.

>From the point of view of measurements, it turns out that realizing an
ITP-meter remains a major challenge, mostly due to the difficulties with
the choice of the dynamics at the contact. Indeed, one would need to find
a dynamics for the contact that satisfies the condition for equalization of
ITPs (that is, the coarse-grained detailed balance relation (\ref{effect-DB})
or a generalization of it),
without having
information on the probability weights of the system on which the
measure should be done.
Alternatively, measurements of ITPs using the variance of the globally 
conserved quantity in subsystems of ``mesoscopic'' size seem to be a promising route.

\section*{Acknowledgments}

This work has been partly supported by the Swiss National Science Foundation.

\appendix

\section{Independence of the choice of subsystems}
\label{app-choice-indep}

In this appendix, we show that, according to the claim made in
Sec.~\ref{sec-frame-def}, the value of the ITP $\lambda_k$ does not
depend on the partition chosen. To this aim, let us relate
the ITP $\lambda_k$ obtained from the subsystems, to a global ITP
defined on the whole system. 
Knowing that 
\begin{equation}\label{dist-norm}
\int \prod_{k=1}^\ell dQ_{ka}\; \Psi(\{Q_{ka}\}|\{Q_k\}) = 1
\end{equation}
and neglecting the correction term $\epsilon_N$ appearing in
Eq.~(\ref{afp2}) in the large N limit,
one can derive an expression for $Z(\{Q_k\})$ depending on the partition
functions of the subsystems
\begin{equation}\label{part-func-I}
Z(\{Q_k\}) = \int \prod_{k=1}^\ell dQ_{ka}\; Z_a(\{Q_{ka}\})Z_b(\{Q_k-Q_{ka}\})\;.
\end{equation}
Consistently with the additivity property (\ref{afp}), (\ref{afp2}),
we assume a general scaling form for the partition functions $Z_{\nu}$ for
a large number of sites $N_{\nu}$
\begin{equation}\label{scal}
Z_{\nu}(\{Q_{k\nu}\})=A_{\nu} \exp\left\{N_{\nu}\zeta_{\nu}\left(\frac{Q_{1\nu}}{N_{\nu}},\ldots,
\frac{Q_{\ell\nu}}{N_{\nu}}\right)\right\}
\end{equation}
This amounts to assuming that $\ln Z_{\nu}(\{Q_{k\nu}\})$ is extensive.
This choice allows us to perform a saddlepoint approximation for the
integrals in Eq.~(\ref{part-func-I}) and thus we obtain
\begin{eqnarray}\label{part-func}
Z(\{Q_k\})&=&N^\ell Z_a(\{Q^*_{ka}\}) Z_b(\{Q_k-Q^*_{ka}\})\times\nonumber\\
&&\int_0^{q_k}\prod_{k=1}^\ell dq_{ka}\;e^{-N (f_a(q_{ka})+f_b(q_{ka}))}
\end{eqnarray}
where $q_k=Q_k/N$,  $q_{k\nu}=Q_{k\nu}/N$ and $q_{k\nu}^*=Q_{k\nu}^*/N$,
$\nu\in\{a,b\}$. The functions $f_{\nu}$ read with $n_{\nu}=N_{\nu}/N$ 
\begin{eqnarray}
f_a(\{q_{ka}\})&=&-\frac{n_a}{2 }\sum_{i,j=1}^{\ell}\frac{\partial^2\zeta_a}{\partial q_{ia}\partial q_{ja}}\left(\left\{\frac{q_{ka}}{n_a}\right\}\right)\Big\vert_{(q_{ia}^*,q_{ja}^*)}\times\nonumber\\
&&\qquad(q_{ia}-q_{ia}^*)(q_{ja}-q_{ja}^*)\nonumber\\
f_b(\{q_{ka}\})&=&-\frac{n_b}{2 }\sum_{i,j=1}^{\ell}\frac{\partial^2\zeta_b}{\partial q_{ia}\partial q_{ja}}\left(\left\{\frac{q_k-q_{ka}}{n_b}\right\}\right)\Big\vert_{(q_{ia}^*,q_{ja}^*)}\times\nonumber\\
&&\qquad(q_{ia}-q_{ia}^*)(q_{ja}-q_{ja}^*)
\end{eqnarray}
Taking the logarithm of the above expression, Eq.~(\ref{part-func}), leads to
\begin{eqnarray}
\ln Z(\{Q_k\})&=&\ln Z_a(\{Q_{ka}^*\})+\ln Z_b(\{Q_k-Q_{ka}^*\})\nonumber\\
&+&\ln \xi(q_k)
\end{eqnarray}
where $\xi(q_k)$ is given by the finite integrals
of Eq.~(\ref{part-func}). We thus obtain for the derivative with
respect to $Q_k$, using Eq.~(\ref{def-ITP}),
\begin{equation}
\frac{\partial \ln Z(\{Q_k\})}{\partial Q_k}=\lambda_k\frac{\partial  Q_{ka}^*}{\partial Q_k}+\lambda_k\left(1-\frac{\partial Q_{ka}^*}{\partial Q_k}\right)+\frac{1}{N}\frac{\partial \ln \xi}{\partial q_k}\;.
\end{equation}
Since $\xi(q_k)$ is expected to be bounded when $N\to\infty$
while the ratio $q_k=Q_k/N$ is kept constant,
the last term vanishes in the thermodynamic limit, and we find
\begin{equation}
\frac{\partial \ln Z(\{Q_k\})}{\partial Q_k}=\lambda_k
\end{equation}
where the value of $\lambda_k$ stays finite in the limit defined above
due to the scaling form of $Z_{\nu}$ in Eq.~(\ref{scal}).
The fact that we can compute the value of $\lambda_k$ directly from the
whole system shows the independence of the partition chosen.
This means that the
ITPs $\lambda_k$ ($k=1,\dots,\ell$) are global values that can be used
to characterize the macroscopic state of the whole system.

\section{Additivity condition for matrix product ansatz}
\label{app-matrix}

The aim of this appendix is to show that, under reasonable assumptions,
a system described by a matrix product ansatz with
\emph{finite} matrices fulfills the additivity property (\ref{afp}), (\ref{afp2}).
For the sake of simplicity, we only deal with the case of a single conserved
quantity, but generalizations to several conserved quantities are
rather straightforward.
We also use periodic boundary conditions, though calculations
with open boundaries would essentially be similar.
Considering a generic lattice model as defined in Eq.~(\ref{model-MPA}),
the conditional distribution $\Psi(Q_a|Q)$ reads:
\be
\Psi(Q_a|Q) = \frac{1}{Z(Q)} \, \mathrm{Tr} [R_a(Q_a) R_b(Q-Q_a)]
\ee
To proceed further, one needs to introduce the Laplace transforms
$\hat{R}(s)$ and $\hat{M}(s)$ of $R(Q)$ and $M(q)$ respectively, which
leads to $\hat{R}(s)=\hat{M}(s)^N$.
Making the further hypothesis that $\hat{M}(s)$ is invertible, one
can find a matrix $B(s)$ such that $\hat{M}(s)=\exp[B(s)]$,
yielding $\hat{R}(s)=\exp[NB(s)]$.
The matrix $B(s)$ can be decomposed into a sum of two complex-valued matrices
\cite{Horn}:
\be \label{matrix-decomp}
B(s)=D(s)+L(s)
\ee
where $D(s)$ is a diagonalizable matrix, and $L(s)$ is a nilpotent one,
meaning that there exists a positive integer $p(s)$ such that $L(s)^{p(s)}=0$.
In addition, the matrices $D(s)$ and $L(s)$ commute.
Using Eq.~(\ref{matrix-decomp}), one has
\be
\hat{R}(s) = e^{ND(s)} \, e^{NL(s)}\\
= e^{ND(s)} \, {\cal P}_s(NL(s))
\ee
where ${\cal P}_s$ is a polynomial of degree $p(s)-1$, since higher order
terms in the expansion of $\exp[NL(s)]$ vanish.
Accordingly, in the large $N$ limit, the dominant contribution to
$\hat{R}(s)$ is proportional to $N^{p(s)-1} \exp[ND(s)]$.
Then, as $D(s)$ is diagonalizable, there exists a matrix $V(s)$ such that
$D(s)=V^{-1}(s) \Lambda(s) V(s)$, where $\Lambda(s)$ is the diagonal matrix. 
It results that
\be
e^{ND(s)} = V^{-1}(s) \, e^{N\Lambda(s)} V(s)
\ee
so that the dominant contribution to $\exp[ND(s)]$ is proportional to
$\exp[N\ell_1(s)]$, $\ell_1(s)$ being the eigenvalue of $D(s)$ with
the largest real part. Altogether, $\hat{R}(s)$ takes 
the following form, to leading order in $N$:
\be
\hat{R}(s) \sim N^{p(s)-1} \exp(N\ell_1(s)) K(s)
\ee
where $K(s)$ is a matrix that does not depend on $N$.
Then $R(Q)$ is obtained through an inverse Laplace transform:
\be
R(Q) \sim \int_{s_0-i\infty}^{s_0+i\infty} \frac{ds}{2\pi i}\,
N^{p(s)-1} \exp[N(\ell_1(s)+s\rho)] K(s)
\ee
with $\rho = Q/N$, and where $s_0$ is an arbitrary real number,
greater than the real part of
all the singularities of the integrand. Assuming that the equation
$G(s)=d\ell_1/ds+\rho=0$ has a solution $s^*$, and that $s_0$ can
be chosen as $s^*$, a saddle-point calculation
shows that the dominant contribution to $R(Q)$ is proportional to
$\exp[N(\ell_1(s^*)+s^*\rho)]$. 
The remaining Gaussian integral around the saddle (on the imaginary axis)
converges since $G''(s)=\ell_1''(s)>0$. This last property can be shown
in the following way. The logarithm of the grand-canonical partition
function satisfies $\ln \tilde{Z}(s) \approx N\ell_1(s)$, with the
identification $\lambda = s$. Then, from Eq.~(\ref{cum}),
the second derivative of $\ln \tilde{Z}$
is the variance of $Q$, which is positive; hence $\ell_1''(s)>0$.
Since $Z(Q)=\mathrm{Tr} R(Q)$, $Z(Q)$ is also proportional to
$\exp[N(\ell_1(s^*)+s^*\rho)]$. As a result, one finds
\bea
\ln \mathrm{Tr} [R_a(Q_a) R_b(Q_b)] &\approx&
N_a(\ell_1^a(s_a^*)+s_a^*\rho_a) \\ \nonumber
&& \qquad \qquad + N_b(\ell_1^b(s_b^*)+s_b^*\rho_b) \\ \nonumber
&\approx& \ln Z_a(Q_a) + \ln Z_b(Q_b)\; ,
\eea
from which the additivity condition (\ref{afp}), (\ref{afp2}) follows.

\section{Pair factorized steady state for a continuous mass model}
\label{app-pair}

Let us show that the model introduced in Sec.~\ref{pair}, defined through
the transport rate given in Eq.~(\ref{pair-rate}), 
leads to a pair factorized steady state (\ref{pair-dist}).
This can be verified with the master equation for this model, which
reads:
\begin{widetext}
\begin{eqnarray}
\frac{dP(\{m_i\},t)}{dt}&=&\sum_{i=1}^N \int_0^{m_{i+1}}d\mu\,
\vp(\mu|m_{i-1},m_i+\mu, m_{i+1}-\mu) P(\ldots m_i+\mu,m_{i+1}-\mu,\ldots,t)
\nonumber\\
&-&\sum_{i=1}^N \int_0^{\infty}d\mu\,\vp(\mu|m_{i-1},m_i, m_{i+1})
P(\{m_i\},t)\;
\end{eqnarray}
\end{widetext}
where $P(\{m_i\},t)$ describes the time evolution of the probability of a
microstate $\{m_i\}$. In the steady state the time derivative vanishes.
Plugging Eqs.~(\ref{pair-rate}) and (\ref{pair-dist}) into the master
equation and dividing by $Z^{-1} \prod_{i=1}^N g(m_i,m_{i+1})$ yields:
\begin{eqnarray} \nonumber
&&\sum_{i=1}^N\int_0^{m_{i+1}}d\mu\, v(\mu)\frac{g(m_i,m_{i+1}-\mu)}{g(m_i,m_{i+1})}\frac{g(m_{i+1}-\mu,m_{i+2})}{g(m_{i+1},m_{i+2})}\\
&=&\sum_{i=1}^N\int_0^{\infty}d\mu\, v(\mu)\frac{g(m_{i-1},m_i-\mu)}{g(m_{i-1},m_i)}\frac{g(m_i-\mu,m_{i+1})}{g(m_i,m_{i+1})}
\end{eqnarray}
Substituting $i+1$ by $i$ on the left hand side and knowing that $g(m,n)=0$ for
values of $m$ or $n$ smaller than zero one finds the above equality verified.

\end{document}